\documentclass[
reprint,
superscriptaddress,
showpacs,
amsmath,amssymb,
aps,
prb,
floatfix
]{revtex4-1}

\usepackage{graphicx}
\usepackage{dcolumn}
\usepackage{bm}

\begin{document}
	
	
	\title{Microwave induced tunable subharmonic steps in superconductor-ferromagnet-superconductor Josephson junction }
	
	\author{M. Nashaat}
	\email{
		majed@sci.cu.edu.eg}
	\affiliation{%
		Department of Physics, Cairo University, Cairo, 12613, Egypt
	}%
	\affiliation{%
		BLTP, JINR, Dubna, Moscow Region, 141980, Russian Federation
	}%
	
	\author{Yu. M. Shukrinov}
	\email{
		shukrinv@theor.jinr.ru}
	\affiliation{%
		BLTP, JINR, Dubna, Moscow Region, 141980, Russian Federation
	}%
	\affiliation{%
		Dubna State University, Dubna, 141982, Russian Federation
	}%

	\author{A. Irie}
	\affiliation{%
		Department of Electrical and Electronic Systems Engineering, Utsunomiya University, Utsunomiya, Japan.
	}%
	\author{A.Y. Ellithi}
	\affiliation{%
		Department of Physics, Cairo University, Cairo, 12613, Egypt
	}%
	\author{Th. M.  El Sherbini}
	\affiliation{%
		Department of Physics, Cairo University, Cairo, 12613, Egypt
	}%

	\begin{abstract}
		We investigate the coupling between ferromagnet  and superconducting phase dynamics in superconductor-ferromagnet-superconductor Josephson junction. The current-voltage characteristics of the junction demonstrate a pattern of subharmonic current steps which forms a devil's staircase structure. We show that a width of the steps becomes maximal at ferromagnetic resonance. Moreover, we demonstrate that the structure of the steps and their widths can be tuned by changing the frequency of the external magnetic field, ratio of Josephson to magnetic energy, Gilbert damping and the junction size.
	\end{abstract}

	\maketitle
	
\textbf{		This paper is submitted to LTP Journal.}
	\section{Introduction}
	
	Josephson junction with ferromagnet layer (F) is widely considered to be the place where spintronics and superconductivity fields interact\cite{Linder2014}. In these junctions the supercurrent induces magnetization dynamics due to the coupling between the Josephson and magnetic subsystems.  The possibility of achieving electric control over the magnetic properties of the magnet via Josephson current and its counterpart, i.e., achieving magnetic control over Josephson current, recently attracted a lot of attention \cite{Linder2014,Shukrinovepls2018,shukrinov2017magnetization,buzdin2008direct,Buzdin2005,Bergeret2005,Golubov2004}. The current-phase relation in the superconductor-ferromagnet-superconductor junction  (SFS) junctions is very sensitive to the mutual orientation of the magnetizations in the F-layer\cite{ silaev2017anomalous, bobkova2017gauge}. In Ref.[\onlinecite{Shukrinov2019}] the authors demonstrate a unique magnetization dynamics with a series of specific phase trajectories. The origin of these trajectories is related to a direct coupling between the magnetic moment and the Josephson oscillations in these junctions. 
	
	External electromagnetic field can also provide a coupling between spin wave and Josephson phase in SFS junctions \cite{weides20060,pfeiffer2008static,hikino2011ferromagnetic,wild2010josephson,kemmler2010magnetic,volkov2009hybridization,mai2011interaction}. Spin waves are elementary spin excitations which considered to be as both spatial and time dependent variations in the magnetization \cite{Maekawa2009,ounadjela2003spin}.  The ferromagnetic resonance (FMR) corresponds to the  uniform precession of the magnetization around an  external applied magnetic field \cite{Maekawa2009}. This mode can be resonantly excited by ac magnetic field that couples directly to the magnetization dynamics as described by the Landau-Lifshitz-Gilbert (LLG) equation \cite{Maekawa2009,ounadjela2003spin}.

	In Ref.[\onlinecite{Maekawa2009}] the authors show that spin wave resonance at frequency $\omega_{r}$ in SFS implies a dissipation that is manifested as a depression in the IV-characteristic of the junction when $\hbar\omega_{r} = 2eV$, where $\hbar$ is the Planck's constant, e is the electron charge and $V$ is the voltage across the junction. The ac Josephson current produces an oscillating magnetic field and when the Josephson frequency matches the spin wave frequency, this resonantly excites the magnetization dynamics $M(t)$ \cite{Maekawa2009}. Due to the nonlinearity of the Josephson effect, there is a rectification of current across the junction, resulting in a dip in the average dc component of the suppercurrent \cite{Maekawa2009}.
	
	In Ref.[\onlinecite{hikino2011ferromagnetic}] the authors neglect the effective field due to Josephson energy  in LLG equation and the results reveal that even steps appear in the  IV-characteristic of SFS junction under external magnetic field. The origin of these steps is due to the interaction of Cooper pairs with even number of magnons.  Inside the ferromagnet, if the Cooper pairs scattered by odd number of magnons, no Josephson current flows due to the formation of spin triplet state \cite{hikino2011ferromagnetic}. However,  if the Cooper pairs interact with even number of magnons, the Josephson coupling between the s-wave superconductor is achieved and the spin singlet state is formed, resulting in flows of Josephson current\cite{hikino2011ferromagnetic}. In Ref.[\onlinecite{shukrinov2018}] we show that taking into account the effective field due to Josepshon energy and at FMR, additional subharmonic current steps appear in the IV-characteristic for overdamped SFS junction with spin wave excitations (magnons). It is found that the position of the current steps in the IV-characteristics form devil's staircase structure which follows continued fraction formula \cite{shukrinov2018}. The positions of those fractional steps are given by
	\begin{equation}
	V=\left( N \pm \frac{1}{n\pm
		\frac{1}{m\pm\frac{1}{p\pm..}}}\right)\Omega,
	\label{eq8}
	\end{equation}
	where $\Omega=\omega/\omega_c$, $\omega$ is the frequency of the external radiation, $\omega_c$ is the is the characteristic frequency of the Josephson junction  and $N$, $n$, $m$, $p$ are positive integers.

	In this paper, we present a detailed analysis for the IV-characteristics of SFS junction under external magnetic field, and show how we can control the position of the subharmonic steps and alter their widths. The coupling between spin wave and Josephson phase in SFS junction is achieved through the Josephson energy and gauge invariant phase difference between the S-layers. In the framework of our approach, the dynamics of the SFS junction is fully described by the resistively shunted junction  (RSJ) model and LLG equation. These equations are solved numerically by the $4^{th}$ order Runge-Kutta method. The appearance and position of the observed current steps depend directly on the magnetic field and junction parameters.

	\section{Model and methods}
	\begin{figure}[!ht]
		\centering
		\includegraphics[width=0.5\linewidth, angle =0]{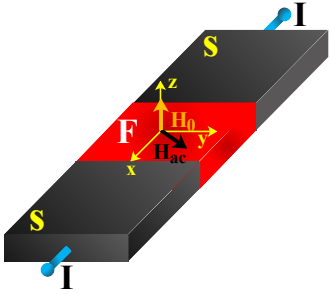}
		\caption [Sketch for SFS Josephson junction with spin wave excitations] {SFS Josephson junction. The bias current is applied in x-direction, an  external magnetic field with amplitude $H_{ac}$ and frequency $\omega$ is applied in xy-plane and an uniaxial constant magnetic field $H_0$ is applied in z-direction.
		}
		\label{fig:SFS}
	\end{figure}
	In Fig \ref{fig:SFS} we consider a current biased SFS junction where the two superconductors are separated by ferromagnet layer with thickness $d$. The area of the junction is $ L_{y}L_{z}$. An uniaxial constant magnetic field $H_0$ is applied in z-direction, while the magnetic field is applied in xy-plane $ \mathbf{H}_{ac}= (H_{ac} \cos \omega t, H_{ac} \sin \omega t, 0 )$ with amplitude $H_{ac}$ and frequency $\omega$. The magnetic field is induced in the F-layer through $\mathbf{B}(t)=4\pi \mathbf{M}(t)$, and the magnetic fluxes in z- and y-direction are  $\Phi_{z}(t)=4 \pi d L_{y} M_{z}(t)$, $\Phi_{y}(t)=4 \pi d L_{z} M_{y}(t)$, respectively. The gauge-invariant phase difference in the junction is given by \cite{likharevbook}:
	\begin{eqnarray}
	\nabla_{y,z} \theta(y,z,t) = - \frac{2 \pi d}{\Phi_0} \textbf{B}(t) \times \textbf{n},
	\end{eqnarray}
	where $\theta$ is the phase difference between superconducting electrodes, and  $\Phi_{0}=h/2e$ is the magnetic flux quantum and $\textbf{n}$ is a unit vector normal to yz-plane. The gauge-invariant phase difference in terms of magnetization components reads as
		\begin{eqnarray}	
	\theta (y,z,t)=\theta (t)-\frac{ 8 \pi^{2}  d
		M_z(t)}{\Phi _{0}} y+\frac{8 \pi^{2}  d M_y(t)}{\Phi _{0}} z,
	\end{eqnarray}
	where	$\Phi_{0}=h/(2e)$ is the magnetic flux quantum.

According to RSJ model, the current through the junction is given by \cite{hikino2011ferromagnetic}:
\begin{eqnarray}
\frac{I}{I^{0}_{c} } &=&\sin \theta(y,z,t) + \frac{\Phi_{0}}{2\pi I^{0}_{c}  R} \frac{d\theta(y,z,t)}{dt}, 
\label{eq:RSJj}
\end{eqnarray}
	where   $I^{0}_{c}$ is the critical current, and R is the resistance in the Josephson junction. After taking into account the gauge invariance including the magnetization of the ferromagnet and integrating over the junction area the electric current reads \cite{hikino2011ferromagnetic}:
	\begin{eqnarray}
	\frac{I}{I^{0}_{c} } &=& \frac{\Phi _o^2 \sin (\theta(t)) \sin \left(\frac{4 \pi^{2}  d M_{z}(t)  L_{y}}{\Phi _o}\right) \sin \left(\frac{4 \pi^{2}  d M_{y}(t) L_{z}}{\Phi _o}\right)}{16 \pi ^4 d^2 L_{z}L_{y}M_{z}(t)M_{y}(t)}\nonumber \\ &+&  \frac{\Phi_{0}}{2\pi R I^{0}_{c}} \frac{d\theta(y,z,t)}{dt}.
	\label{equ:RCSJ2}
	\end{eqnarray}

	The applied magnetic field in the xy-plane causes precessional motion of the magnetization in the F-layer. The dynamics of magnetization $\textbf{M}$ in the F-layer is described by  LLG equation
	{\small
		\begin{eqnarray}
		(1+\alpha^{2}) \frac{d\textbf{M}}{dt} = -\gamma \ \textbf{M} \times \textbf{H}_{eff} - \frac{\gamma \ \alpha}{| \textbf{M}|} \left[ \textbf{M} \times (\textbf{M} \times \textbf{H}_{eff})\right] \
		\label{eq:LLG}
		\end{eqnarray}}
	The total energy of junction in the proposed model is given by $E=E_{s}+E_{M}+E_{ac}$ where  $E_{s}$ is the energy stored in Josephson junction, $E_{M}$ is the energy of uniaxial dc magnetic field (Zeeman energy) and  $E_{ac}$ is the energy of ac magnetic field:
	\begin{eqnarray}
	E_{s\;} &=& -\frac{\Phi_0}{2\pi} \theta (y,z,t) I   + E_{J} \left[1-\cos\left(y,z,t\right)\right],  \nonumber \\
	E_{M} &=& - V_{F} H_{0}  M_{z}(t), \nonumber  \\
	E_{ac}&=& - V_{F} M_{x}(t) H_{ac} \cos (\omega t) - V_{F} M_{y}(t) H_{ac} \sin (\omega t)\
	\label{eq3}
	\end{eqnarray}
	Here, $E_{J}=\Phi_{0} I^{0}_{c}/2\pi$ is the the Josephson energy, $H_{0}=\omega_{0}/\gamma$, $\omega_{0}$ is the FMR frequency, and $V_{F}$ is the volume of the ferromagnet. 	We neglect the anisotropy energy due to demagnetizing effect for simplicity. The effective field in LLG equation is calculated by
	\begin{eqnarray}
	\textbf{H}_{eff}=-\frac{1}{V_{F}}  \nabla_{M} E
	\end{eqnarray}
	Thus, the effective field $\textbf{H}_{m}$ due to microwave radiation $\textbf{H}_{ac}$ and uniaxial magnetic field $\textbf{H}_{0}$ is given by
	\begin{eqnarray}
	\textbf{H}_{m}=H_{ac} \cos(\omega t) \ \hat{\textbf{e}}_{x} +H_{ac} \sin(\omega t) \ \hat{\textbf{e}}_{y} + H_{0} \ \hat{\textbf{e}}_{z}. 
	\end{eqnarray}
	while the effective field ($H_{s}$) due to superconducting part is found from
	\begin{eqnarray}
	\textbf{H}_{s} =-\frac{E_{J} }{V_{F}}\sin(\theta(y,z,t))  \nabla_{M} \theta(y,z,t).
	\end{eqnarray}
	One should take the integration of LLG on coordinates, however, the superconducting part is the only part which depends on the coordinate so, we can integrate the effective field due to the Josephson energy and insert the result into LLG equation. Then, the  y- and z-component are given by
	\begin{eqnarray}
	\textbf{H}_{sy} &=&\frac{E_{J} \cos (\theta(t)) \sin \left(  \pi \Phi_{z}(t)/\Phi_{0}\right)}{V_{F} \pi M_{y}(t)  \Phi_{z}(t)} \bigg[ \Phi_{0}  \cos (\pi\Phi_{y}(t)/\Phi_{0})\nonumber \\ &-&\Phi_{0}^{2}\frac{\sin(\pi\Phi_{y}(t)/\Phi_{0})}{\pi \Phi_{y}(t)} \bigg] \hat{\textbf{e}}_{y},\\
	\textbf{H}_{sz} &=&\frac{E_{J}\cos (\theta(t)) \sin \left(  \pi \Phi_{y}(t)/\Phi_{0}\right)}{V_{F} \pi M_{z}(t)  \Phi_{y}(t)}  \bigg[ \Phi_{0} \cos (\pi\Phi_{z}(t)/\Phi_{0})\nonumber \\ &-&\Phi_{0}^{2}\frac{\sin(\pi\Phi_{z}(t)/\Phi_{0})}{\pi \Phi_{z}(t)} \bigg]  \hat{\textbf{e}}_{z}.
	\end{eqnarray}
	As a result, the total effective field is $\textbf{H}_{eff}=\textbf{H}_{m}+\textbf{H}_{s}$.	
	In the dimensionless form we use  $t\rightarrow t \omega_{c}$, $\omega_{c}=2 \pi I^{0}_{c} R/\Phi_{0}$ is the characteristic frequency,  $\textbf{m}=\textbf{M}/M_{0}$, $M_{0}=\|\textbf{M}\|$, $\textbf{h}_{eff} = \textbf{H}_{eff}/H_{0}$, $\epsilon_{J} =E_{J}/V_{F} M_{0} H_{0}$, $h_{ac}=H_{ac}/H_{0}$,  $\Omega=\omega /\omega_{c}$,  $\Omega_{0}=\omega_{0} /\omega_{c}$, $\phi_{sy}$=$4\pi^2 L_{y} d M_{0}/\Phi_o$, $\phi_{sz}$=$4\pi^2 l_{z} d M_{0}/\Phi_o$.  Finally, the voltage $V(t)=d\theta/dt$ is normalized to $\hbar \omega_c/(2e)$. The LLG and the effective field equations take the form
	\begin{eqnarray}
	\frac{d\textbf{m}}{dt} &=& - \frac{\Omega_{0}}{(1+\alpha^{2})}\bigg(\textbf{m} \times \textbf{h}_{eff} + \alpha \left[ \textbf{m} \times (\textbf{m} \times \textbf{h}_{eff})\right] \bigg) \label{eq:llg_normalized} \nonumber \\
	\end{eqnarray}
	with \begin{eqnarray}
	\textbf{h}_{eff}&=&h_{ac} \cos (\Omega t) \hat{\textbf{e}}_{x} + \left(  h_{ac} \ sin (\Omega t) + \Gamma_{ij} \epsilon_{J}\cos \theta \right)  \hat{\textbf{e}}_{y} \nonumber \\&+& \left( 1 + \Gamma_{ji} \epsilon_{J}\cos \theta \right)  \hat{\textbf{e}}_{z}, \label{eq:hefftotal} \\
	\Gamma_{ij}&=&\frac{ \sin \left(   \phi_{si} m_{j}\right)}{ m_{i} ( \phi_{si} m_{j})} \left[   \cos ( \phi_{sj} m_{i})-\frac{\sin(\phi_{sj} m_{i})}{(\phi_{sj} m_{i})} \right],
	\end{eqnarray}
	where \textit{i}=y, \textit{j}=z. The RSJ in the dimensionless form is given by
	\begin{eqnarray}
	I/I^{0}_{c} = \frac{ \sin \left(\phi_{sy} m_{z}\right)\sin \left(\phi_{sz}m_{y}\right)}{(\phi_{sy}m_{z})(\phi_{sz}m_{y})}  \sin \theta + \frac{d\theta}{dt}.
	\label{equ:theta2}
	\end{eqnarray}
	
	The magnetization and phase dynamics of the SFS junction can be described by solving Eq.(\ref{equ:theta2}) together with Eq.(\ref{eq:llg_normalized}). To solve this system of equations, we employ the fourth-order Runge-Kutta scheme. At each current step, we find the temporal dependence of the voltage $V(t)$,  phase $\theta(t)$, and $m_{i}$ (\textit{i}=x,y,z) in the $(0, T_{max})$ interval. Then the time-average voltage\textit{ V} is given by $V =\frac{1}{T_{f}-T_{i}} \int V (t)dt$, where $T_{i}$ and $T_{f}$ determine the interval for the temporal averaging. The current value is increased or decreased by a small amount of $\delta$I (the bias current step) to calculate the voltage at the next point of the IV-characteristics. The phase, voltage and magnetization components achieved at the previous current step are used as the initial conditions for the next current step.  The one-loop IV-characteristic is obtained by sweeping the bias current from \textit{I}= 0 to \textit{I}= 3 and back down to \textit{I}= 0. The initial conditions for the magnetization components are assumed to be $m_{x}=0$, $m_{y}=0.01$ and $m_{z}=\sqrt{1-m^{2}_{x}-m^{2}_{y}}$, while for the voltage and phase we have $V_{ini}=0$, $\theta_{ini}$=$0$. The numerical parameters (if not mentioned) are taken as $\alpha=0.1$, $h_{ac}=1$, $\phi_{sy}=\phi_{sz}=4$, $\epsilon_{J}=0.2$ and $\Omega_{0}=0.5$.

	\section{Results and discussions}
	It is well-known that Josephson  oscillations can be synchronized by external microwave radiation which leads to Shapiro steps in the IV-characteristic \cite{shapiro1963josephson}. The position of the Shapiro step is determined by relation $V=\frac{n}{m} \Omega$, where \textit{n,m} are integers. The steps at $m=1$ are called harmonics, otherwise we deal with synchronized subharmonic (fractional) steps. We show below the appearance of subharmonics in our case.
	
\begin{figure}[h!]
		\centering
		\includegraphics[width=0.9\linewidth, angle =0]{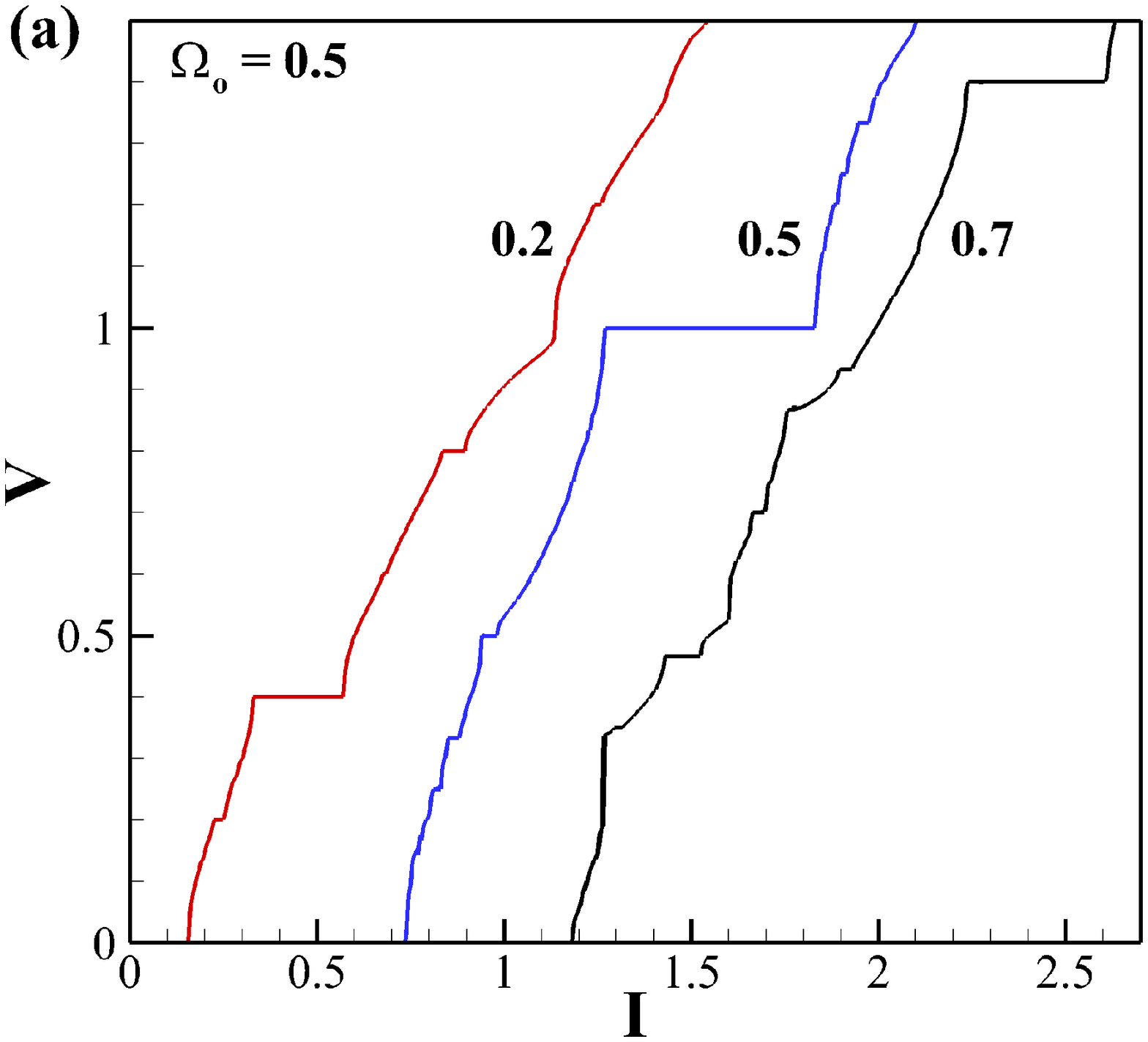}
		\includegraphics[width=0.9\linewidth, angle =0]{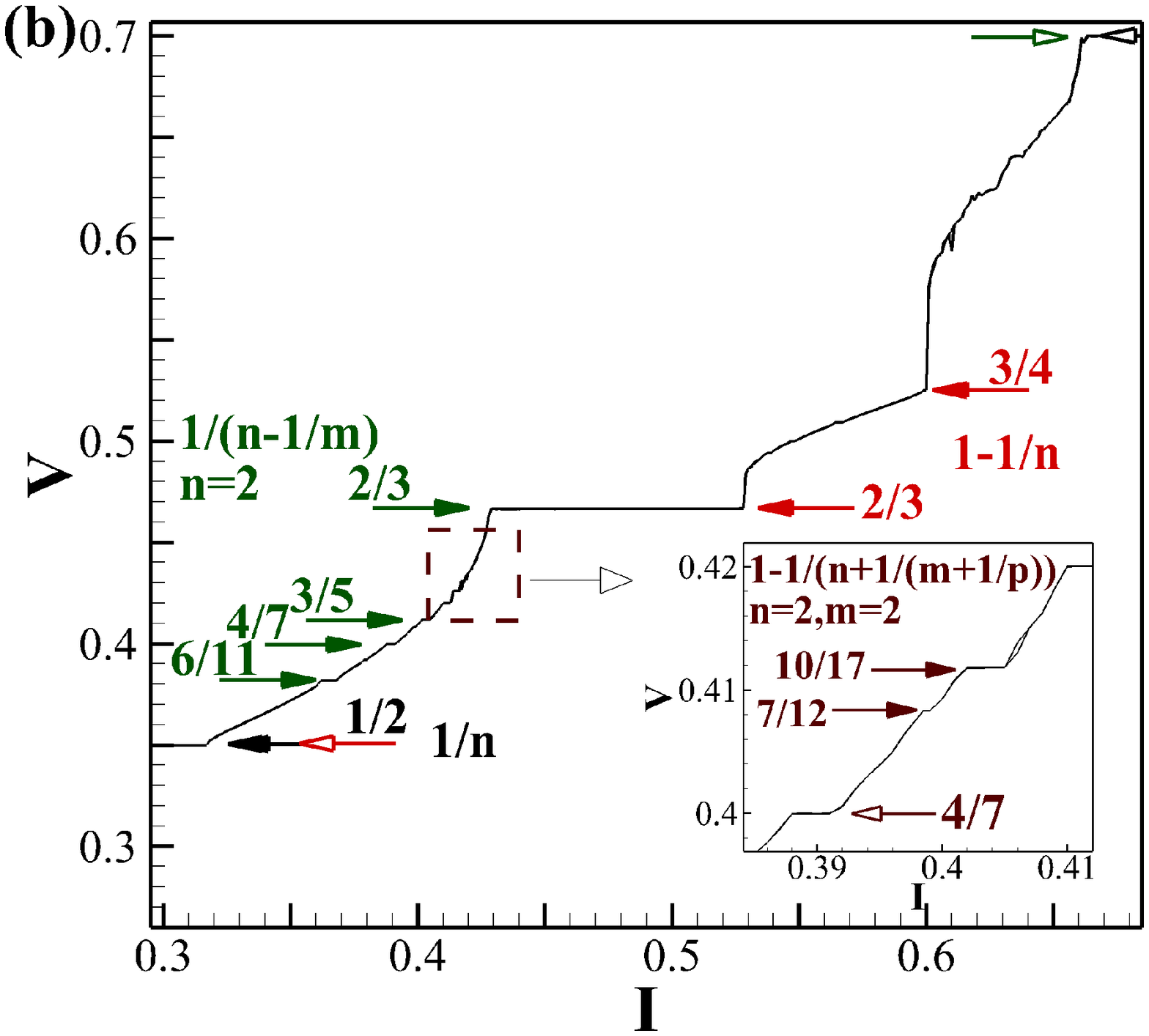}
		\caption {(a) IV-characteristic at three different values of $\Omega$. For clarity, the IV-characteristics for $\Omega=0.5$ and $\Omega=0.7$ have been shifted to the right, by $\Delta I= 0.5$ and $\Delta I= 1$, respectively with respect to $\Omega=0.2$; (b) An enlarged part of the IV-characteristic with $\Omega=0.7$. To get step voltage multiply the corresponding fraction with $\Omega=0.7$. }
		\label{fig:2}
	\end{figure}

	First we present the simulated IV-characteristics at different frequencies of the magnetic field. The IV-characteristics at three different values of $\Omega$ are shown in Fig \ref{fig:2}(a). 
	
	As we see,  the second harmonic has the largest step width at the ferromagnetic resonance frequency $\Omega=\Omega_{0}$, i.e., the FMR is manifested itself by the step's width. There are also many  subharmonic current steps in the IV-characteristic. We have analyzed the steps position between $V=0$ and $V=0.7$ for $\Omega=0.7$ and found different level continued fractions, which follow the formula given by Eq.(\ref{eq8}) and demonstrated in Fig.\ref{fig:2}(b). We see the reflection of the second level continued fractions $1/n$ and $1-1/n$ with $N=1$. In addition to this, steps with third level continued fractions $1/(n-1/m)$ with $N=1$  is manifested. In the inset we demonstrate part of the fourth level continued fraction $1-1/(n+1/(m+1/p))$ with $n=2$ and $m=2$.
	
	In case of external electromagnetic field which leads to the additional electric  current $I_{ac}=A\sin\Omega t$,  the width of the Shapiro step is proportional to  $\propto J_{n}(A/\Omega)$, where $J_{n}$ is the Bessel function of first kind. The preliminary results (not presented here) show that the width of the Shapiro-like steps under external magnetic field has a more complex frequency dependence \cite{shukrinov2018}. This question will be discussed in detail somewhere else.
	
\begin{figure}[h!]
		\centering
		\includegraphics[width=0.9\linewidth, angle =0]{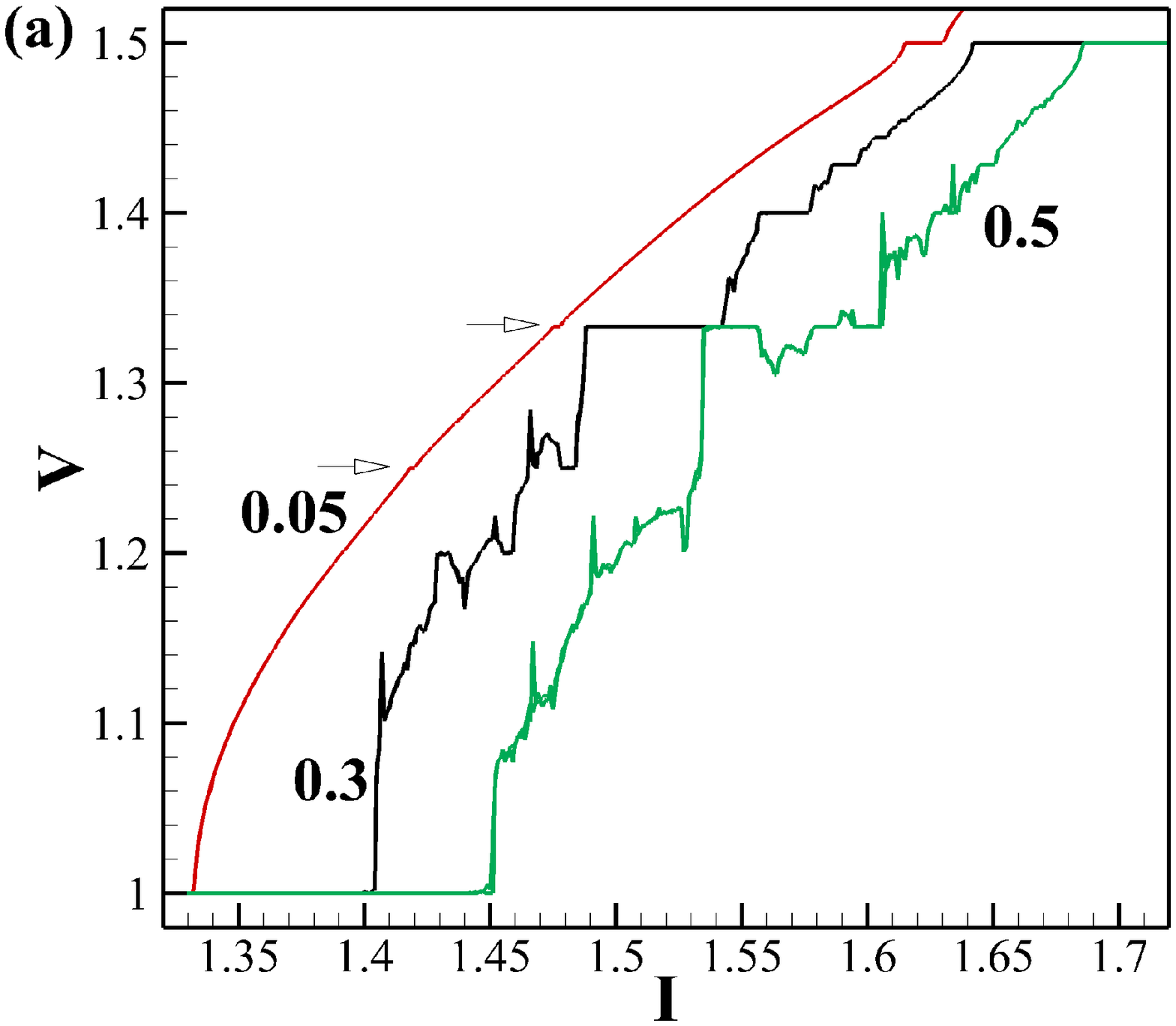}
		\includegraphics[width=0.9\linewidth, angle =0]{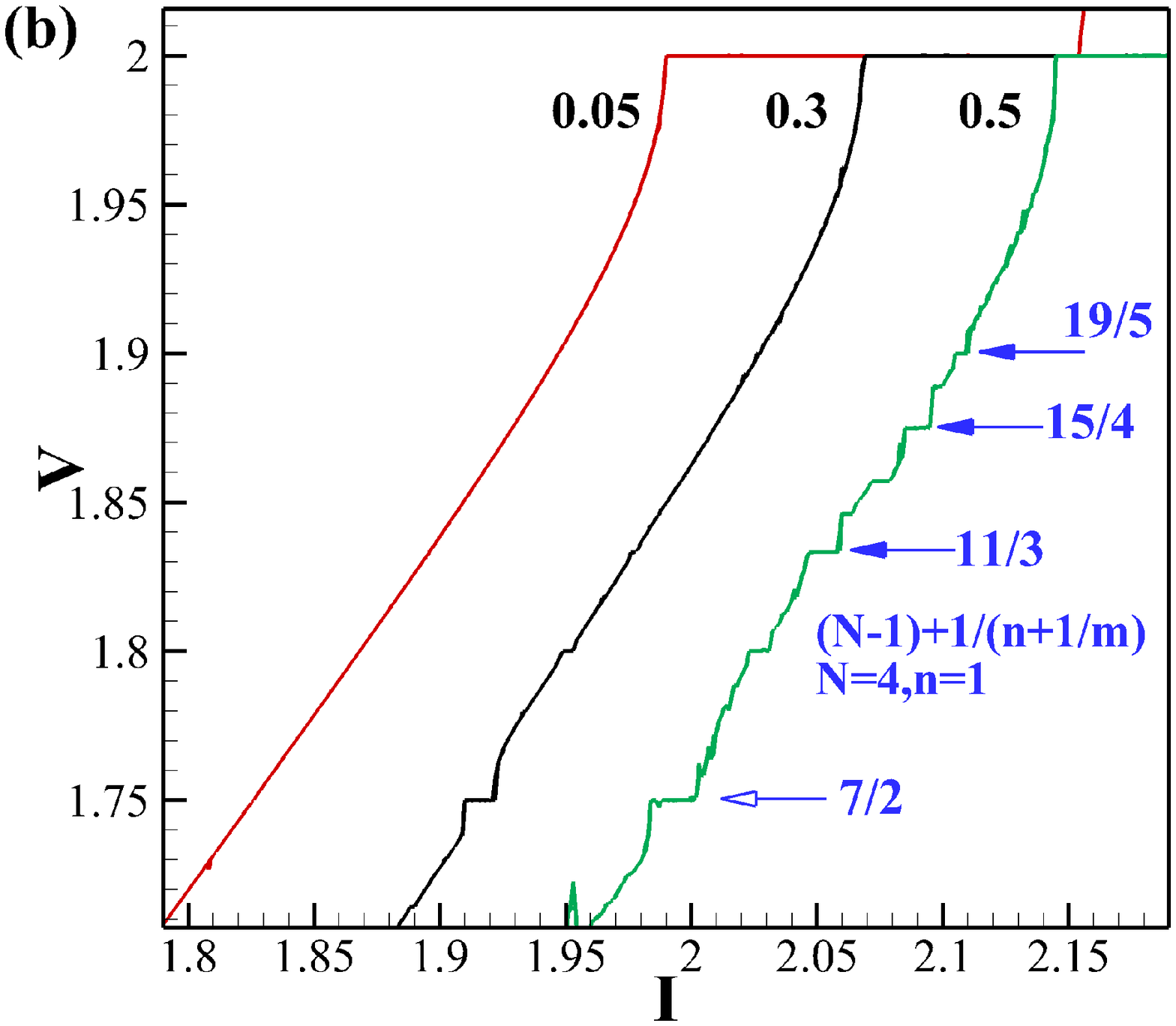}
		\caption { (a) An enlarged part of the IV-characteristic at different values of $\epsilon_{J}$ in the interval between $V=1$ and $V=1.5$; (b) The same in the interval between $V=1.75$ and $V=2$. For clarity, the IV-characteristics for $\epsilon_{J}=0.3$, and $0.5$ have been shifted to right, by $\Delta I= 0.07$, and $0.14$, respectively with respect to the case with $\epsilon_{J}=0.05$. }
		\label{fig:3}
	\end{figure}

	The coupling between Josephson phase and magnetization manifests itself in the appearance of the Shapiro steps in the IV-characteristics at fractional and odd multiplies of $\Omega$ \cite{shukrinov2018}. In Fig.\ref{fig:3} we demonstrate the effect of the ratio of the Josephson to magnetic energy   $\epsilon_{J}$ on appearance of the steps  and their width for $
	\Omega=0.5$ where the enlarged parts of the IV-characteristics at three different values of $\epsilon_{J}$ are shown. As it is demonstrated in the figures, at $\epsilon_{J}=0.05$ only two subharmonic steps appear between $V=1$ and $V=1.5$ (see hollow arrows). An enhanced staircase structure appears by increasing the value of $\epsilon_{J}$ , which can be see at $\epsilon_{J}=0.3$ and $0.5$. Moreover, an intense subharmonic steps appear between $V=1.75$ and $V=2$ for $\epsilon_{J}=0.5$. The positions for these steps reflect third level continued fraction $(N-1)+1/(n+1/m)$ with \textit{N}=4 and \textit{n}=1 [see Fig.\ref{fig:3}(b)]. 

\begin{figure}[h!]
		\centering
		\includegraphics[width=0.9\linewidth, angle =0]{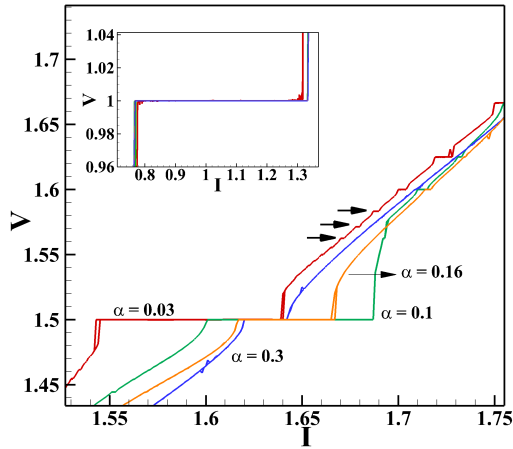}
		\caption {  An enlarged part of IV-characteristic for four different values of Gilbert damping for $\Omega=0.5$. The inset shows an enlarged part of current step with constant voltage at $V=2\Omega$.}
		\label{fig:4}
	\end{figure}

	Let us now demonstrate the effect of Gilbert damping on the devil's staircase structure. The  Gilbert damping $\alpha$ is introduced into LLG equation \cite{gilbert2004,Hickey2009} to describe the relaxation of magnetization dynamics.  To reflect effect of Gilbert damping, we show an enlarged part of the IV-characteristic at three different values of $\alpha$ in Fig.\ref{fig:4}.
	
	The width of current step at $V=2\Omega$ is almost the same at different values of $\alpha$ (e.g., see upward inset $V=2\Omega$). The subharmonic current step width for $V=(n/m)\Omega$ ($n$ is odd, $m$ is integer) is decreasing with increasing $\alpha$. In addition a horizontal shift for the current steps occurs. We see the intense current steps in the IV-characteristic  for small value of $\alpha=0.03$ (see black solid arrows). With increase in Gilbert damping (see $\alpha=0.1$, $0.16$ and $0.3$) the higher level subharmonic steps disappear. It is well-known that at large value of $\alpha$ the FMR linewidth become more broadening and the resonance frequency is shifted from $\Omega_{0}$. Accordingly, the subharmonic steps disappear at large value of $\alpha$. Furthermore, using the formula presented in Ref.[\onlinecite{shukrinov2018}] the width at $\Omega=\Omega_{0}$ for the fractional and odd current steps is proportional to  $(4\alpha^{2}+\alpha^{4})^{-q/2}$ $\times$ $(12+3\alpha^{2})^{-k/2}$, where \textit{q} and \textit{k} are integers.

\begin{figure}[h!]
		\centering
		\includegraphics[width=.9\linewidth, angle =0]{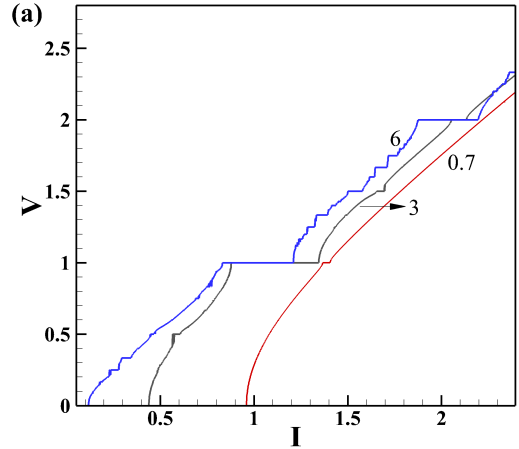}
		\includegraphics[width=.9\linewidth, angle =0]{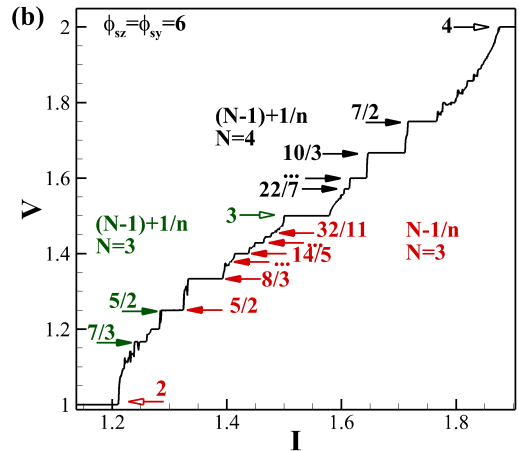}
		\caption { (a) IV-characteristic at three different values of $\phi_{sz}=0.7,3,6$ and $\phi_{sy}=\phi_{sz}$. (b) An enlarged part of the IV-characteristic at $\phi_{sz}=\phi_{sy}$=6. The hollow arrows represent the starting point of the sequences. To get step voltage we multiply the corresponding fraction by $\Omega=0.5$. }
		\label{fig:5}
	\end{figure}

	Finally, we demonstrate the effect of the junction size on the devil's staircase in the IV-characteristic under external magnetic field. The junction size changes the value of $\phi_{sy}$ and $\phi_{sz}$. In Fig.\ref{fig:5}(a) we demonstrate the effect of the junction thickness by changing $\phi_{sz}$ ($\phi_{sy}$ is qualitatively the same).

	We observe an enhanced subharmonic structure with increase of junction size or the thickness of the ferromagnet. In Ref.[\onlinecite{hikino2011ferromagnetic}] the authors demonstrated that the critical current and the width of the step at $V=2\Omega$ as a function of $L_{z}/L_{y}$ follow Bessel function of first kind. In Fig.\ref{fig:5}(b), we can see the  parts of continued fraction sequences for subharmonic steps between $V=1$ and $V=2$ at $\phi_{sz}=\phi_{sy}=6$. Current steps between $V=1$ and $V=1.5$ reflect the two second level continued fractions $(N-1)+1/n$ and $N-1/n$ with $N=3$ in both cases, while for the steps between $V=1.5$ and $V=2$ follow the second level continued fraction $(N-1)+1/n$ with $N=4$.
	
	Finally, we discuss the possibility of experimentally observing the effects presented in this paper.  For junction size $d=5\,{\rm nm}$, $L_{y}=L_{z}=80 \,{\rm nm}$, critical current $I^{0}_{c}\approx200 \,\mu {\rm A}$, saturation magnetization $M_{0} \approx 5\times 10^{5} \,{\rm A}/{\rm m}$,  $H_{0}\approx40\,{\rm mT}$ and gyromagnetic ratio $\gamma=3\pi \,{\rm MHz}/{\rm T}$, we find the value of $\phi_{sy(z)}$=$4\pi^2 L_{y(z)} d M_{0}/\Phi_{0}$ = $4.8$ and $\epsilon_{J}=0.1$. With the same junction parameters one can control the appearance of the subharmonic steps by tuning the strength of the constant magnetic field $H_{0}$. Estimations show that for $H_{0}=10 \,{\rm mT}$, the value of $\epsilon_{J}=0.4$, and the fractional subharmonic steps are enhanced. In general, the subharmonic steps are sensitive to junction parameters, Gilbert damping and the frequency of the external magnetic field.

	\section{Conclusions}
	In this work, we have studied the IV-characteristics of superconductor-ferromagnet-superconductor Josephson junction under external magnetic field. We used a modified RSJ model which hosts magnetization dynamics in F-layer. Due to the external magnetic field, the coupling between magnetic moment and Josephson phase is achieved through the effective field taking into account the Josephson energy and gauge invariant phase difference between the superconducting electrodes. We have solved a system of equations which describe the dynamics of the Josephson phase by the RSJ equation and magnetization dynamics by Landau-Lifshitz-Gilbert equation. The IV-characteristic demonstrates subharmonic current steps. The pattern of the subharmonic steps can be controlled by tuning the frequency of the ac magnetic field. We show that by increasing the ratio of the Josephson to magnetic energy an enhanced staircase structure appears. Finally, we demonstrate that Gilbert damping and junction parameters can change the subharmonic step structure. The observed features might find an application in superconducting spintronics.

	\section{Acknowledgment}
	
	We thank Dr. D. V. Kamanin  and Egypt – JINR collaboration for support this work.  The reported study was partially funded by the RFBR research
	Projects No. 18-02-00318 and No. 18-52-45011-IND. Numerical calculations have been made in the framework of the RSF Project No.
	18-71-10095.

\end{document}